\documentclass[superscriptaddress,showpacs,preprintnumbers,amsmath,amssymb,floatfix,elsart,prd]{revtex4}

\usepackage{graphicx}
\usepackage{dcolumn}
\usepackage{bm}
\usepackage{epstopdf}

\newcommand{\beq}{\begin{equation}}
\newcommand{\eeq}{\end{equation}}
\newcommand{\bey}{\begin{eqnarray}}
\newcommand{\eey}{\end{eqnarray}}

\begin{document}

\title{Singularity-free nonexotic compact star in $f(R,T)$ gravity}

\author{Anil Kumar Yadav}
\email{abanilyadav@yahoo.co.in} \affiliation{Department of Physics, United College of Engineering and Research, \\ Greater Noida, India.}

\author{Monimala Mondal}
\email{monimala.mondal88@gmail.com} \affiliation{Department of
Mathematics, Jadavpur University, Kolkata 700032, West Bengal,
India}

\author{Farook Rahaman}
\email{rahaman@iucaa.ernet.in} \affiliation{Department of
Mathematics, Jadavpur University, Kolkata 700032, West Bengal,
India}

\date{\today}

\begin{abstract}
In the present work, we have searched the existence of anisotropic and non-singular compact star in the $f(R,T)$ gravity by taking into account the non-exotic equation of state (EoS). In order to obtain the solutions of the matter content of compact object, we assume the well known barotropic form of EoS that yields the linear relation between pressures and energy density. We propose the existence of non-exotic compact star which shows the validation of energy conditions and stability with in the $f(R,T)$ extended theory of gravity perspective. The linear material correction in the extended theory the matter content of compact star remarkably able to satisfies energy condition. We discuss various physical features of compact star and show that proposed model of stellar object satisfies all regularity conditions and is stable as well as singularity-free.
\end{abstract}

\maketitle

 \section{Introduction}
\label{intro}
Harko et al \citep{Harko/2011} have proposed the extended theory of gravity so called $f(R,T)$ by changing the geometrical part of the Einstein field equations instead of changing the
source side by taking a generalized functional
form of the argument to address galactic, extra-galactic, and
cosmic dynamics. In this theory, the gravitational terms of total action is defined by the functional form of $f(R)$ and $f(T)$. The main aim of this theory is to address some observational phenomenons such as dark energy \citep{Riess/2004}, dark matter \citep{Akerib/2017}, massive pulsars \citep{Antoniadis/2013} that were hardly explain by General Relativity (GR). Among the all extended/modified theories of gravity, the $f(R,T)$ theory attract more due to it's unique feature that is non-minimal coupling of matter and geometry \citep{Yadav/2018a}. In the recent past, several applications of $f(R,T)$ \citep{Moraes/2017,Yadav/2014,Yadav/2018,Moraes/2015,Singh/2014,Shabani/2013,Shabani/2014,Sharif/2014,Reddy/2013} have been reported in the literature. \\

In this paper, we  focus ourselves to investigate a non-exotic compact star with in $f(R,T) = R + 2\zeta T$ formalism where $\zeta$ being the arbitrary constant. The compact star is a hypothetical dense body that may be black hole or degenerate star and the pressure inside it is not isotropic. In Astrophysics, the structure and properties of compact star had been studies by numerous authors in different physical context \citep{Bowers/1974,Herrera/2004,Sharma/2001}.
In 2006, the most significant compact star has been observed by Rosat Surveys due to their X-rays emission \citep{Aqueros/2006}. This means that the gravitational energy of compact star is radiated through X-rays. A long ago, Hewish et al \citep{Hewish/1968} had investigated the some rapidly pulsating radio source which is in general the beam of electromagnetic radiation. This discovery inspire physicists to think about modeling of compact star like neutron star and quark star in framework of general relativity and its extended form \citep{Maurya/2015,Maurya/2016,Maurya/2017,Das/2016}.  {Recently Paul et al \cite{Paul/2019} have studied the cooling of neutron star including axion emission by nucleon - nucleon axion bremsstrahlung}. It is common understanding that one can not analyze the structure and properties of compact star by taking into account of equation of state which relate the pressure and energy density in proportion. In Refs. \cite{Maurya/2016a,Aziz/2016,Rahaman/2014a}, it has been found that the pressure of compact star is anisotropic in nature. In the recent past, Momeni et al \citep{Momeni/2017} have constructed a model of compact star in Horndeski theory of gravity and analyzed it in modified theory of gravity. However, our model deals with the singularity-free compact star composed by non-exotic matter in $f(R,T)$ theory of gravity and its functional form $f(R,T) = R + 2\zeta T$. In Refs. \citep{Moraes/2017,Moraes/2018}, some applications of $f(R,T)$ theory with respect to steller objects are reported. Some other relevant investigations on different functional forms of extended $f(R,T)$ theory of gravitation can be observed in following studies \citep{Zubair/2016a,Alhamazawi/2016} under different physical context. In 2014, Rahaman et al \citep{Rahaman/2014} have studied the static Wormhole in f(R) Gravity with Lorentzian distribution which generates two models - one is derived from power law form and second model is based on assumption of particular shape function which allows the reconstruction of the $f(R)$ theory. Zubair et al \citep{Zubair/2016} have investigated numerical solutions for different wormhole matter content in the realm of $f(R,T)$ gravity. Moraes et al \citep{Moraes/2015} have constructed the model of static Wornhole by applying $f(R,T)$ formalism.\\

In the present paper, we are concerned with the singularity-free nonexotic model of compact star with the realm of functional form of $f(R,T) = R + 2\zeta T$. It is worth to mention that our model is derived from the well known barotropic equation of state (EoS) in Krori and Barua (KB) space-time \cite{Krori/1975} that yield the singularity-free solution. In Ref. \cite{Das/2016}, the authors have investigated a model of stellar object in the static spherically symmetric space-time which is probably singular and generate a set of solutions
describing the interior of a compact star under $f(R,T)$
theory of gravity which admits conformal motion whereas the present investigation is one with singularity-free solution. However a common feature of both the investigations is the non-exotic matter configuration in $f(R,T)$ gravity.\\

The paper is structured as follows: The basics of $f(R,T) = f(R)+2\zeta T$ formalism are presented in section II. Section III deals with the the KB metric, solution of field equations and physical behavior of the model. In section IV, we provide the boundary conditions, which are essential for finding the values of constants.  In section V, we demonstrate the validity of energy conditions, stability and mass -radius relation to show the physical acceptance of model. In Section VI, we match the model parameters with observation data sets. In section VII, we point out our results and discuss the future perspectives of the study.

\section{The $f(R,T) = f(R) + 2\zeta T$ Formalism}
The total action for the $f(R,T)$ theory of gravitation \citep{Harko/2011} reads
\begin{equation}
\label{Harko}
S = \frac{1}{4\pi}\int d^{4}xf(R,T)\sqrt{-g}+\int d^{4}xL_{m}\sqrt{-g}
\end{equation}
where $R$ is the Ricci scalar, $T$ is the trace of energy-momentum tensor $T_{j}^{i}$, $g$ is the metric determinant and $L_{m}$ is the matter Lagrangian density.\\
By varying the total action $S$ with respect to metric $g_{ij}$, we obtain
\begin{equation}
\label{S1}
\begin{array}{ll}
 R_{ij}f^{\prime}(R,T)-\frac{1}{2}f(R,T)g_{ij}+
\left(g_{ij}\bigtriangledown^{i} \bigtriangledown_{i}  - \bigtriangledown_{i} \bigtriangledown_{j}\right)f^{\prime}(R,T)\\
\\
\,\,\,\,\,\,\,
 = 8\pi T_{ij} -\dot{f}(R,T)\theta_{ij}-\dot{f}(R,T)T_{ij}
\end{array}
\end{equation}

Here, $f^{\prime}(R,T) = \frac{\partial f}{\partial R}$ and $\dot{f}(R,T) = \frac{\partial f}{\partial T}$ and $\theta_{ij}$ is read as\\
\begin{equation}
\label{th}
\theta_{ij}=g^{ij}\frac{\partial T_{ij}}{\partial g^{ij}}
\end{equation}
In this paper, we take the more generic form matter Lagrangian as $L_{m} = -\rho$ \cite{Moraes/2017}. Hence equation (\ref{th})leads to
\begin{equation}
\label{th1}
\theta_{ij}=-2T_{ij}-\rho g_{ij}
\end{equation}
Following the proposition of Refs. \citep{Harko/2011}, we assume the functional form of $f(R,T) = f(R)+2\zeta T$ with $\zeta$ is constant. In the literature, this functional form is commonly used to obtain cosmological solution in $f(R,T)$ theory of gravitation \cite{Yadav/2014,Yadav/2018,Moraes/2015}.\\
The equations (\ref{S1}) and (\ref{th1}) lead to
\begin{equation}
\label{G}
G_{ij} = (8\pi+2\zeta)T_{ij}+\zeta(2\rho+T)g_{ij}
\end{equation}
 where $G_{ij}$ is Einstein's tensor.\\

\section{The KB Metric and Field  Equations}
The Krori and Barua space-time \citep{Krori/1975,Rahaman/2012} is read as
\begin{equation}
 \label{spacetime}
ds^{2}=-e^{\nu(r)}dt^{2}+e^{\lambda(r)}dr^{2}+r^{2}(d\theta^{2}+sin^{2}\theta d\phi^{2})
\end{equation}
with $\lambda(r)= Ar^{2}$ and $\nu(r) = Br^{2} + C$ having $A$, $B$ and $C$ are constants.\\
In this paper, we take an anisotropic fluid satisfying the matter content of stellar object as
\begin{equation}
\label{di}
T_{ij} = diag(-\rho, p_{r}, p_{t}, p_{t})
\end{equation}
where $\rho$, $p_{r}$ and $p_{t}$ are the energy density, radial pressure and tangential pressure respectively. Thus the trace of energy momentum tensor may be expressed as $T = -\rho+p_{r}+2p_{t}$.\\

The metric (\ref{spacetime}) and field equation (\ref{G}) along with equation (\ref{di}) lead the following equations:
\begin{equation}
\label{fe1}
 e^{-\lambda}\left(\frac{\lambda^{\prime}}{r}-\frac{1}{r^{2}}\right)+\frac{1}{r^{2}}=(8\pi+\zeta)\rho-\zeta(p_{r}+2p_{t})
\end{equation}
\begin{equation}
\label{fe2}
e^{-\lambda}\left(\frac{\nu^{\prime}}{r}+\frac{1}{r^{2}}\right)-\frac{1}{r^{2}}=\zeta\rho+(8\pi+3\zeta)p_{r}+2\zeta p_{t}
\end{equation}
\begin{equation}
\label{fe3}
 \frac{e^{-\lambda}}{2}\left(\frac{{\nu^{\prime}}^{2}-\lambda^{\prime}\nu^{\prime}}{2}+\frac{\nu^{\prime}-\lambda^{\prime}}{r}+\nu^{\prime\prime}\right)=\zeta\rho+\zeta p_{r}+(8\pi+4\zeta)p_{t}
\end{equation}
\subsection{Solution of Field Equations \& Physical Parameters}
To solve the above set of equations for the matter content of compact star, it is useful to invoke the equation of state (EoS) which gives the relation between energy density and pressure. The most common barotropic forms of EoS \citep{Azreg/2015} are
\begin{equation}
\label{eos1}
p_{r} = \alpha\rho
\end{equation}
\begin{equation}
\label{eos2}
p_{t} = \beta\rho
\end{equation}
where $\alpha$ and $\beta$ are constants having values in the range (0,1).\\

Now, from metric (\ref{spacetime}), one may obtain $\lambda^{\prime} = 2Ar$, $\nu^{\prime} = 2Br$ and $e^{-\lambda} = e^{-Ar^{2}}$. Putting these values in equations (\ref{fe1})$-$(\ref{fe3}) along with the barotropic EoS (\ref{eos1}) and (\ref{eos2}), we obtain
\begin{equation}
\label{rho}
\rho=\frac{1}{[8\pi+\zeta-\zeta(\alpha+2\beta)]}\left[exp(-Ar^{2})\left(2A-\frac{1}{r^{2}}\right)+\frac{1}{r^{2}}\right]
\end{equation}
\begin{equation}
\label{p_{r}}
p_{r}=\frac{\alpha}{[8\pi+\zeta-\zeta(\alpha+2\beta)]}\left[exp(-Ar^{2})\left(2A-\frac{1}{r^{2}}\right)+\frac{1}{r^{2}}\right]
\end{equation}
\begin{equation}
\label{p(t)}
p_{t}=\frac{\beta}{[8\pi+\zeta-\zeta(\alpha+2\beta)]}\left[exp(-Ar^{2})\left(2A-\frac{1}{r^{2}}\right)+\frac{1}{r^{2}}\right]
\end{equation}
We observe that the barotropic EoS (\ref{eos1}) and (\ref{eos2}) are identically satisfied with solutions (\ref{rho})$-$(\ref{p(t)}). Also we note that the energy density $(\rho)$, radial pressure $(p_{r})$ and tangential pressure $(p_{t})$ decrease with r and finally approaches to a small positive values. The behavior of $\rho$, $p_{r}$ and $p_{t}$ is graphed in Fig. 2 for physically acceptable values of problem parameters. Fig. 1 depicts the variation of $\lambda$ against r. \\
\begin{figure*}[thbp]
\begin{center}
\includegraphics[width=0.5\textwidth]{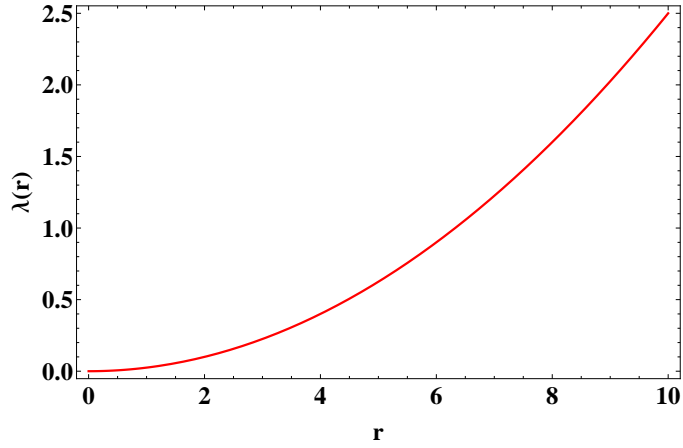}
\caption{Plot of  $\lambda$ versus $r$.}
\label{fig:1.eps}
\end{center}
\end{figure*}
\begin{figure*}[thbp]
\begin{tabular}{rl}
\includegraphics[width=0.5\textwidth]{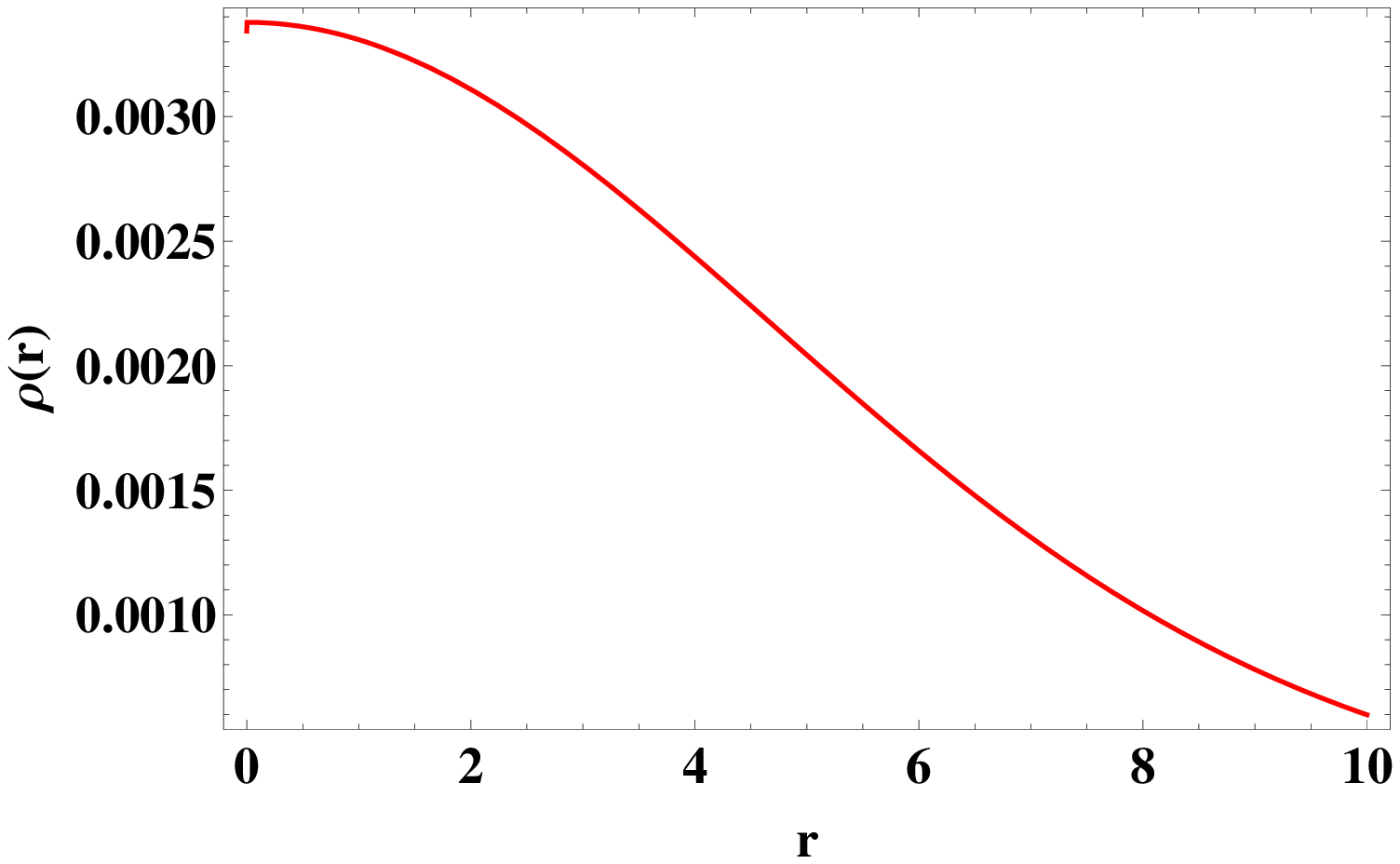}\\
\includegraphics[width=0.5\textwidth]{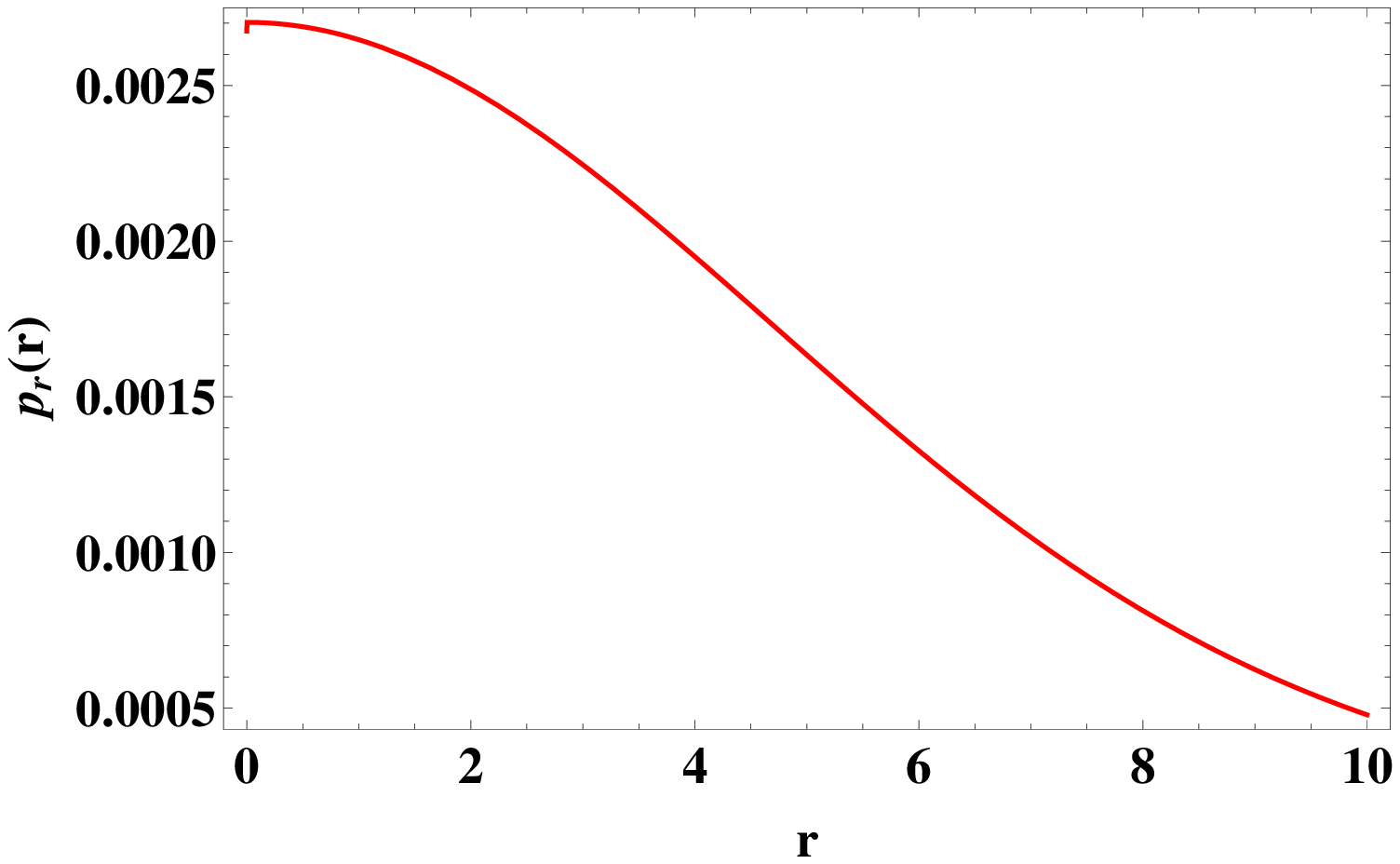}\\
\includegraphics[width=0.5\textwidth]{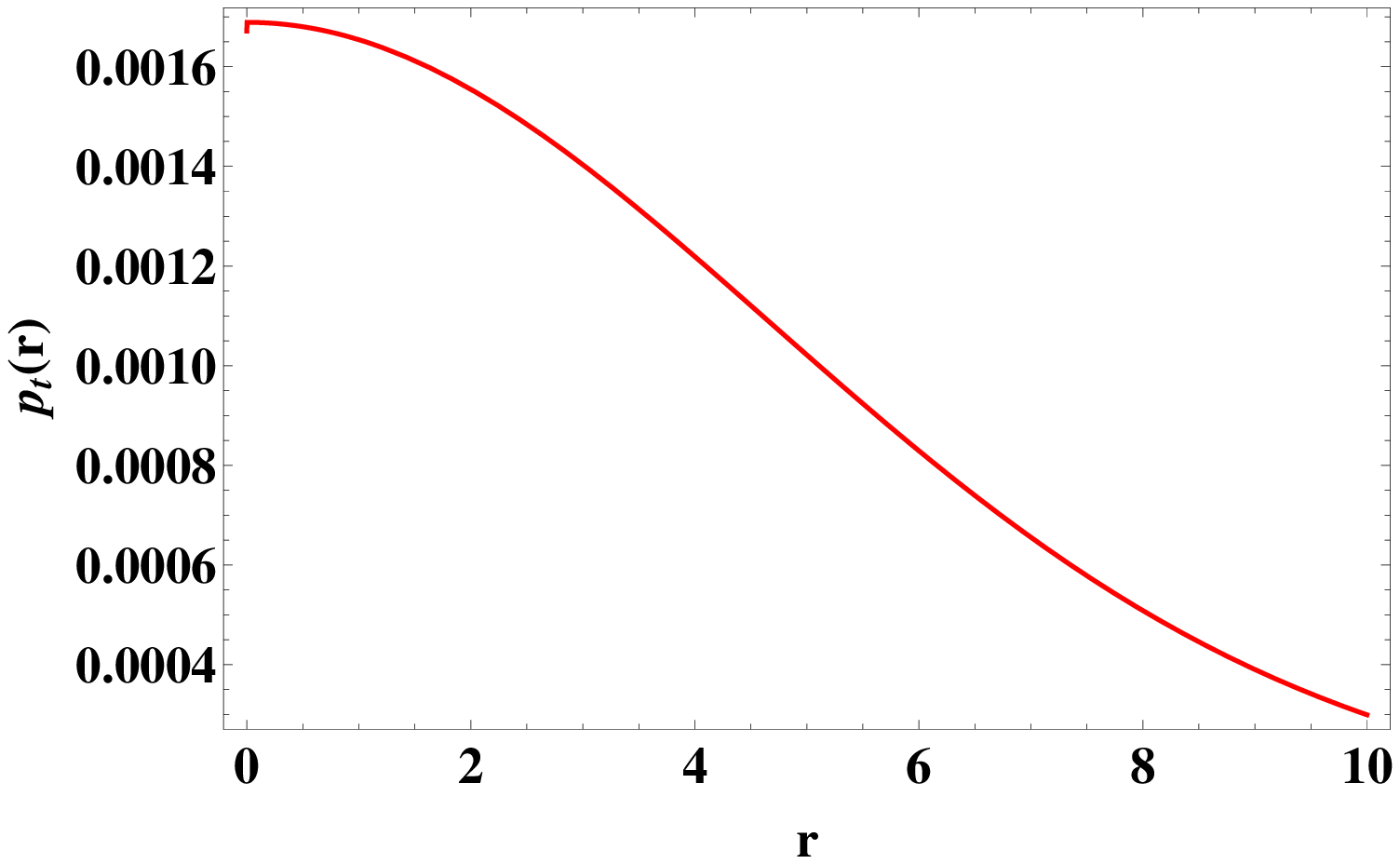}
\end{tabular}
\caption{Plots of $\rho(r)$ vs $r$ (upper panel), $p_{r}(r)$ vs $r$ (middle panel) and $p_{t}(r)$ vs $r$ (lower panel).}
\label{fig:2.pdf}
\end{figure*}

It is worth to note that $\frac{d\rho}{dt}$ and $\frac{dp_{r}}{dt}$ are negative which lead the following requirements for our model to be physically acceptable:\\
(i)~ The energy density is positive and its first derivative is negative.\\
(ii)~ The radial pressure is positive and radial pressure gradient is negative.\\

We also note that at $r = 0$, the second derivative of energy density as well as radial pressure are negative which shows that the energy density and radial pressure are maximum at the center of wormhole. \\

The anisotropic parameter $(\triangle)$ is computed as
\begin{equation}
\label{anisotropy}
\triangle = \frac{2(\beta-\alpha)}{r[8\pi+\zeta-\zeta(\alpha+2\beta)]}\left[exp(-Ar^{2})\left(2A-\frac{1}{r^{2}}\right)+\frac{1}{r^{2}}\right]
\end{equation}

The anisotropic parameters is equivalent to a force due to the local anisotropy which is directed inward if radial pressure is greater than the tangential pressure and outward when radial pressure is less than tangential pressure. From equation (\ref{anisotropy}), we observe that the nature of $\triangle$ depends on the free parameters $\alpha$ and $\beta$. These parameters are positive constant having values in between 0 and 1 but $(\beta - \alpha)$ may be positive or negative depending upon the choice of values of these parameters. Thus the repulsive anisotropic force $(\triangle > 0)$ will appear when $\beta > \alpha$. under this specification, the compact star allows the construction of more massive distribution \cite{Rahaman/2012} that is why we have taken $\beta > \alpha$ throughout the all graphical analysis of the model. Fig. 3 shows the variation of $\triangle$ with respect to r for different choice of $\alpha$ and $\beta$.\\
\begin{figure}[tbp]
\begin{center}
\includegraphics[width=0.5\textwidth]{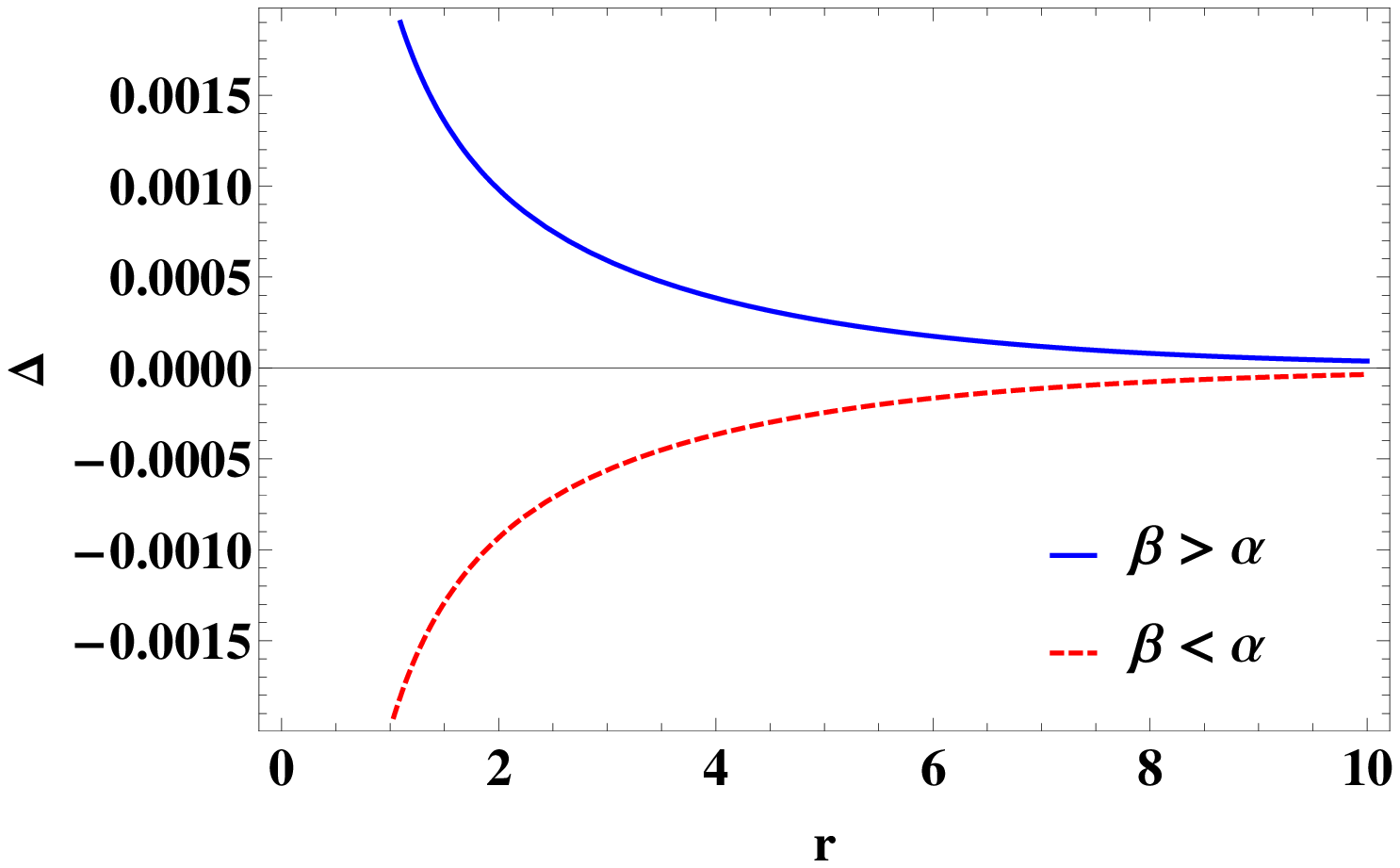}
\caption{The variation of the force $\triangle$ due to the local anisotropy against $r$.}
\label{fig:3.pdf}
\end{center}
\end{figure}
\section{Boundary Conditions}
{
The central density is obtained by putting $r = 0$ in equation (\ref{rho}) i. e.
\begin{equation}
\label{rhoc}
\rho_{c}= \rho(r = 0) = \frac{3A}{8\pi + \zeta(1-\alpha-2\beta)}
\end{equation}
The radial pressure and tangential pressure at center are given by
\begin{equation}
\label{p-r}
p_{rc} = p_{r}(r = 0) = \frac{3A\alpha}{8\pi + \zeta(1-\alpha-2\beta)}
\end{equation}
\begin{equation}
\label{p-t}
p_{tc} = p_{t}(r = 0) = \frac{3A\beta}{8\pi + \zeta(1-\alpha-2\beta)}
\end{equation}
At center anisotropy is zero which leads $\alpha = \beta$. It is also required that the physical fluids must obey the
Zeldovich's criterion i. e. $\frac{p_{rc}}{\rho_{c}}\leq 1$. This implies that $\alpha = \beta \leq 1$. This shows the physical constraints on $\alpha$ and $\beta$.\\
The surface density is obtained by putting $r =R$ in equation (\ref{rho}) $i. e.$
\begin{equation}
\label{rhos}
\rho(r = R) = \frac{\left(1-\frac{2M}{R}\right)\left(2A-\frac{1}{R^{2}}\right)+\frac{1}{R^{2}}}{8\pi +\zeta(1-\alpha-2\beta)}
\end{equation}
To obtain the boundary condition, we will compare the interior metric to the Schwarzschild exterior at the boundary $r = R$ which leads the following equations:
\begin{equation}
\label{be1}
1-\frac{2M}{R} = e^{BR^{2}+C}
\end{equation}
\begin{equation}
\label{be2}
e^{AR^{2}}\left(1-\frac{2M}{R}\right) = 1
\end{equation}
\begin{equation}
\label{be3}
\frac{M}{R^{3}} = Be^{BR^{2}+C}
\end{equation}
The values of constants $A$ and $B$ are evaluated by choosing the boundary conditions such that $p_{r} = 0$ at r = R and $\rho = a = constant$ at r = 0.  Thus, solving equations (\ref{rho}), (\ref{p_{r}}) and (\ref{be1})$-$(\ref{be3}) along with boundary conditions, we obtain
\begin{equation}
\label{A}
A=\frac{[8\pi+\zeta(1-\alpha-2\beta)]a}{3} = \frac{1}{R^{2}}ln\left[1-\frac{2M}{R}\right]^{-1}
\end{equation}
\begin{equation}
\label{B}
B=\frac{1}{2R^{2}}\left[e^{\frac{[8\pi+\zeta(1-\alpha-2\beta)]aR^{2}}{3}}-1\right] = \frac{M}{R^{3}}\left(1-\frac{2M}{R}\right)^{-1}
\end{equation}
From Equation (\ref{A}), it is evident that A is a positive constant and its numerical value can be constraint by the specific choice of other free parameters namely $\zeta$, $\alpha$ and $\beta$. In Refs. \citep{Buchdahl/1959}, Buchdahl has obtained that the maximum allowable compactness for a fluid sphere is $\frac{2M}{R} < 8/9$. In table 1, we have presented the numerical values of model parameters A, B, central density and radial pressure for different strange star candidates. In this paper, we have  chosen $A =0.00541$ for graphical analysis. Applying Buchdahl criteria for compactness in equation (\ref{A}), the chosen value of $A$ gives $R = 9.3749$ (see Table 2) which is very close to observed value of R \citep{Guver/2010a}.\\
}
\begin{table*}
\small
\caption{Determination of model parameters A \& B
for different star candidates} \label{tbl-1}
\begin{tabular}{@{}crrrrrrrrrrr@{}}
\tableline
S. N. & Stars & $A$~~~ &~~~ $B$~~ &~~ $\rho(r =0)(gm/cm^{3})$~~ &~~ $p(r=0)(dyne/cm^{2})$ \\
\tableline

1. & PSRJ 1614-2230 ~&~ $0.00213$~~ &~~ $0.00128$ & $0.361232\times10^{15}$ & $2.682920 \times 10^{35}$    \\
2. & PSRJ 1903+327 ~&~  $0.00489$ ~~&~~ $0.00306$ & $0.815675\times10^{15}$ & $6.058156 \times 10^{35}$   \\
3. & 4U 1820-30 ~&~  $0.00515$ ~~&~~ $0.00321$ & $0.859045\times10^{15}$ & $6.380260 \times 10^{35}$  \\
4. & VelaX-1 ~&~  $0.00506$ ~~&~~ $0.00322$ & $0.841611\times10^{15}$ & $6.250785 \times 10^{35}$ \\
5. & 4U 1608-51 ~&~  $0.00541$ ~~&~~ $0.00345$ & $0.899825\times10^{15}$ & $6.683146 \times 10^{35}$  \\
\tableline
\end{tabular}
\end{table*}
\section{Physical Consequences of Model Under $f(R,T)$ Gravity}
\subsection{Validity of energy conditions}
In this section, we check the validity of energy conditions namely null energy condition (NEC), weak energy condition (WEC), dominant energy condition (DEC) and strong energy condition (SEC) for the proposed compact star. The violation of energy conditions lead the possible cause of existence of exotic matter in compact star. In Refs. \citep{Hochberg/1998}, the EMT violates the NEC at the center (r = 0).\\
(i)NEC:~ $\rho \geq 0$\\
(ii)WEC:~ $\rho + p_{r} \geq 0$ and $\rho + p_{t} \geq 0$\\
(iii)DEC:~ $\rho - p_{r} \geq 0$ and $\rho - p_{t} \geq 0$\\
(iv) SEC:~ $\rho + p_{r} \geq 0$ and $\rho + p_{r} + 2p_{t} \geq 0$\\
\begin{figure*}[thbp]
\begin{tabular}{rl}
\includegraphics[width=0.50\textwidth]{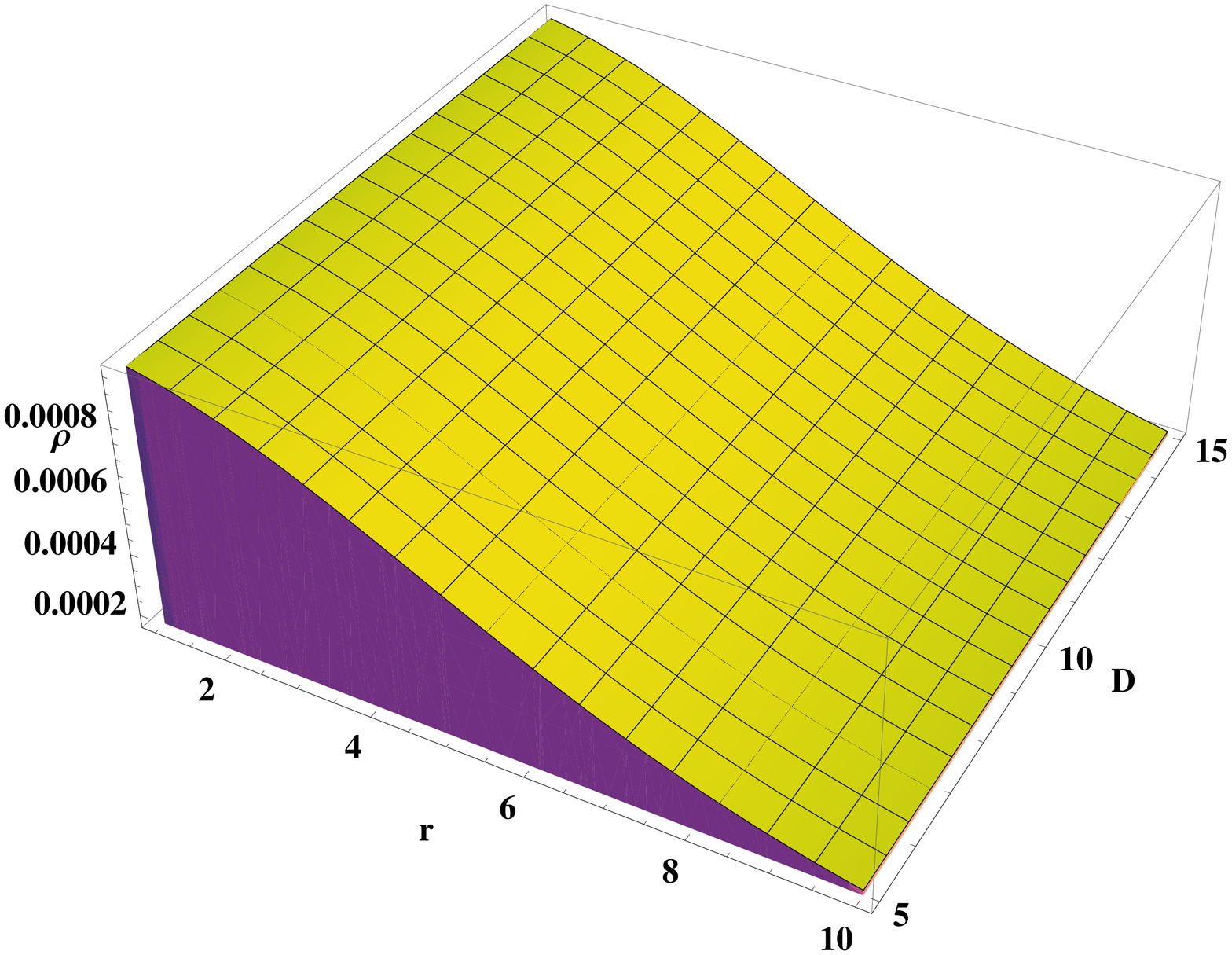}
\includegraphics[width=0.50\textwidth]{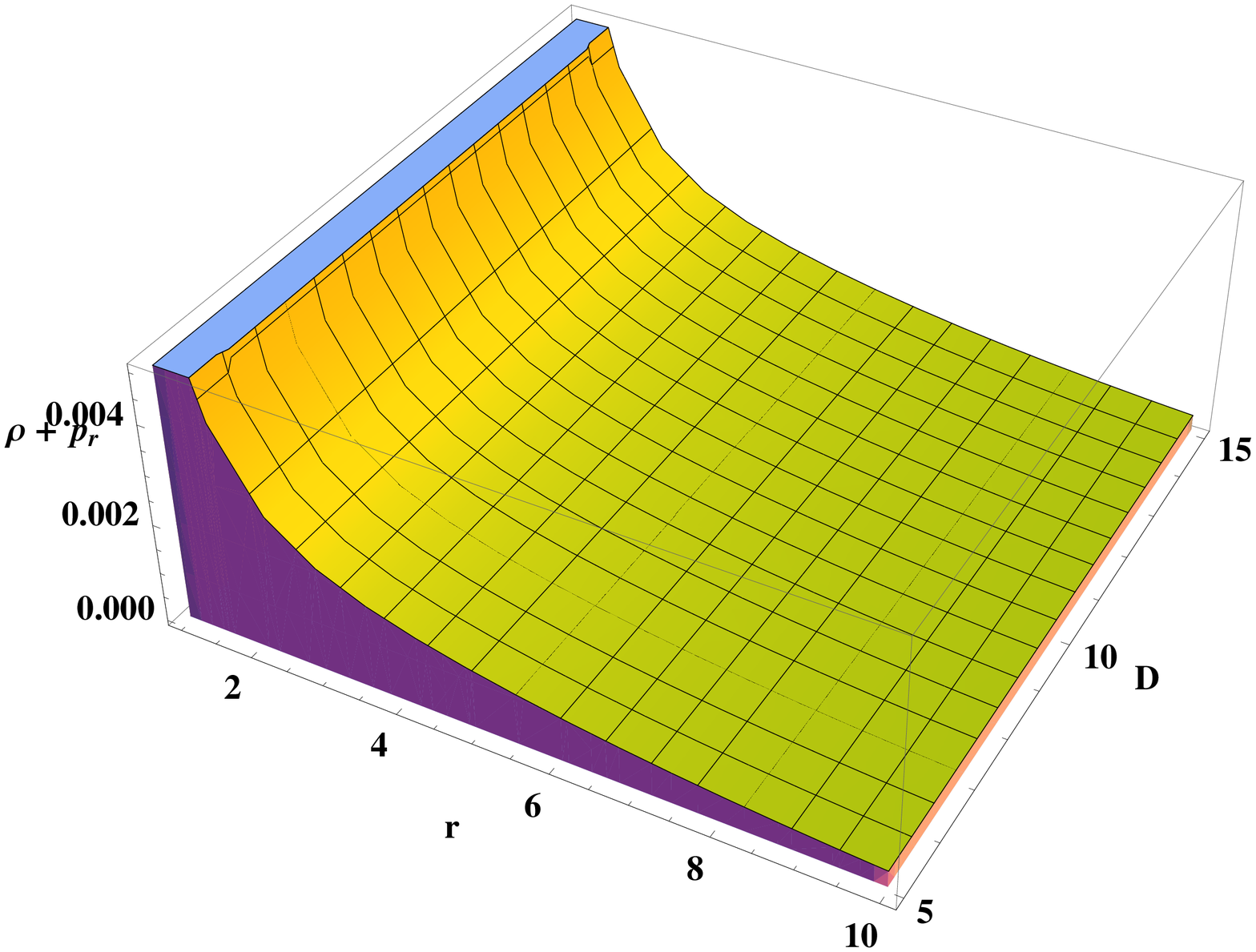}\\
\includegraphics[width=0.50\textwidth]{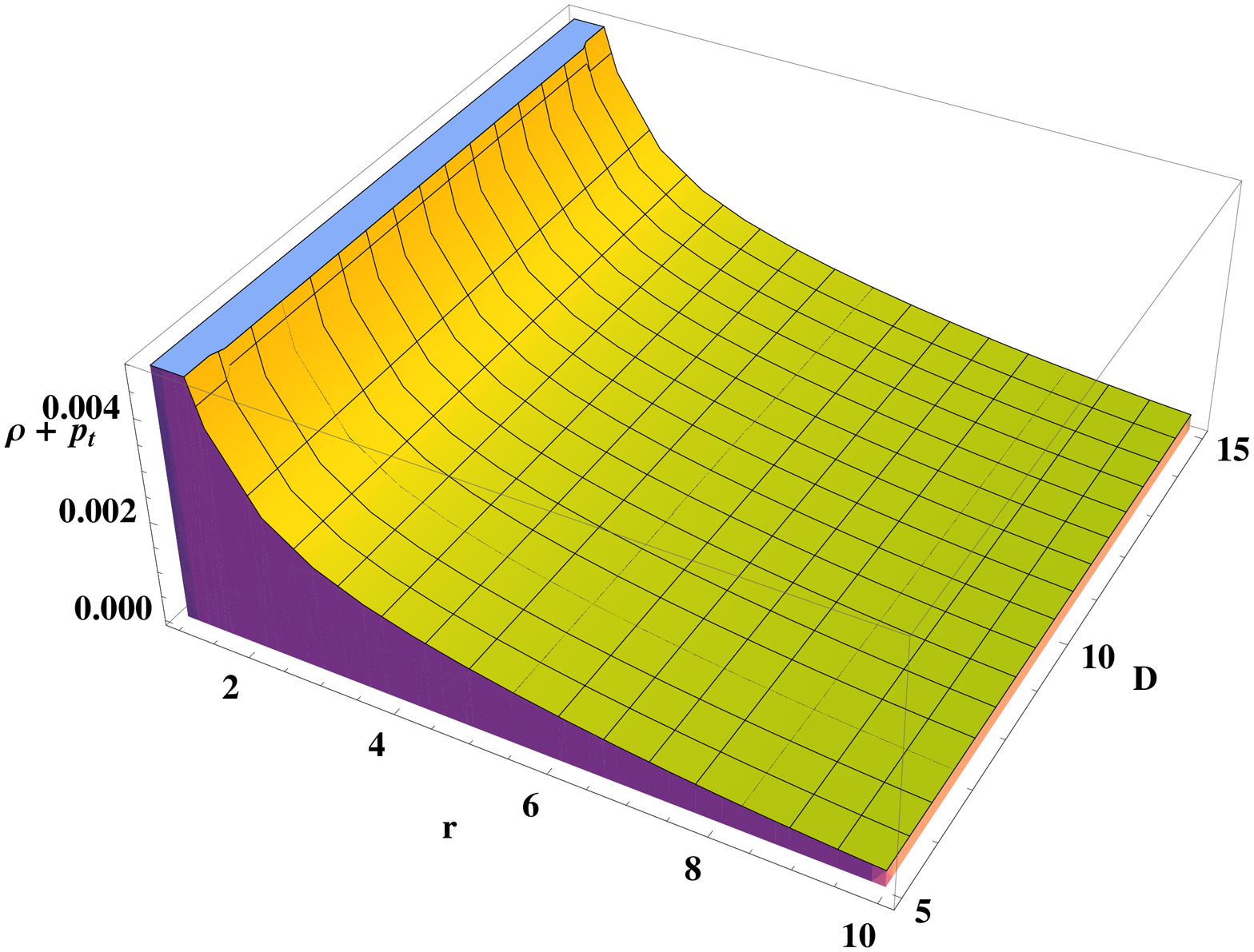}
\includegraphics[width=0.50\textwidth]{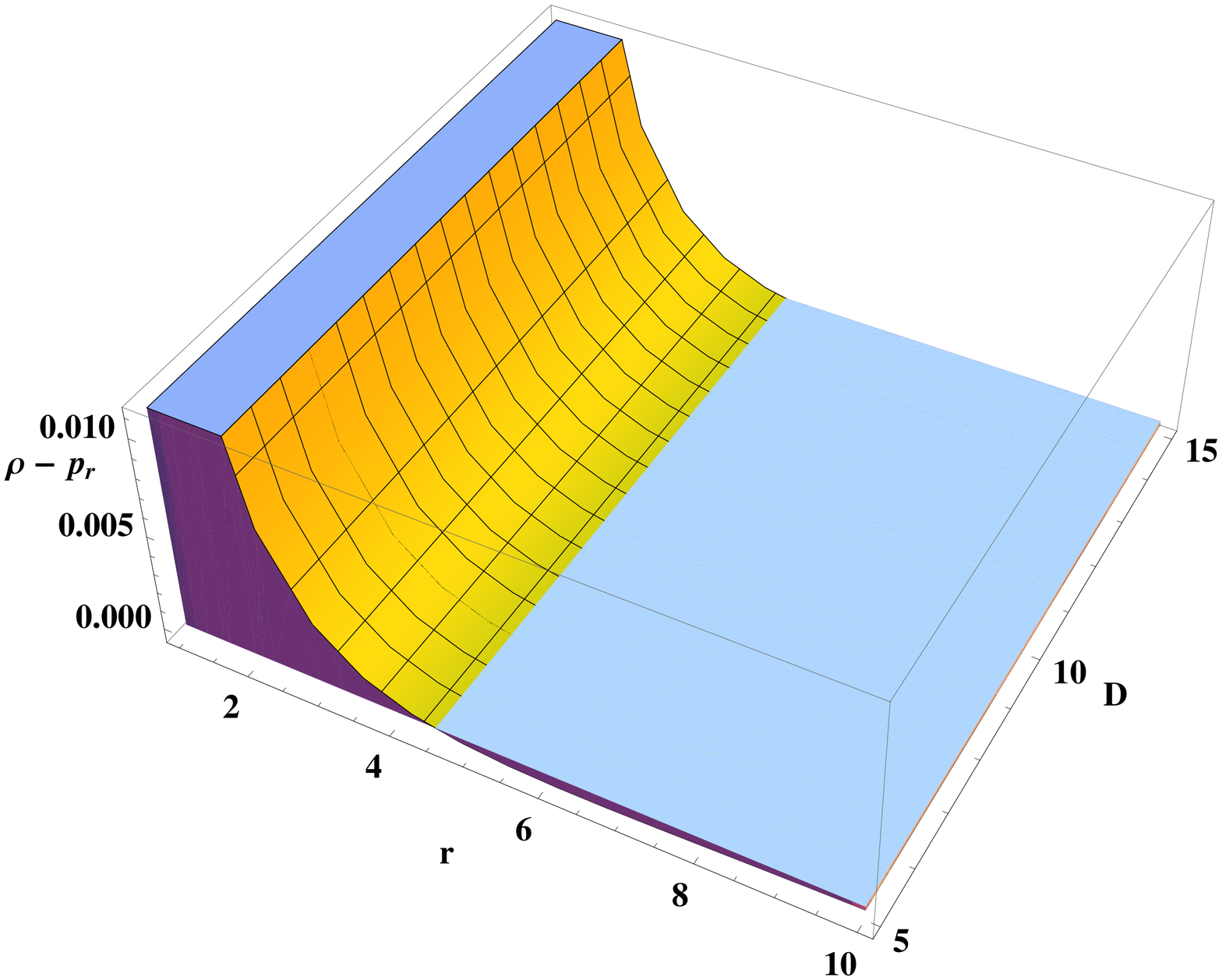}\\
\includegraphics[width=0.50\textwidth]{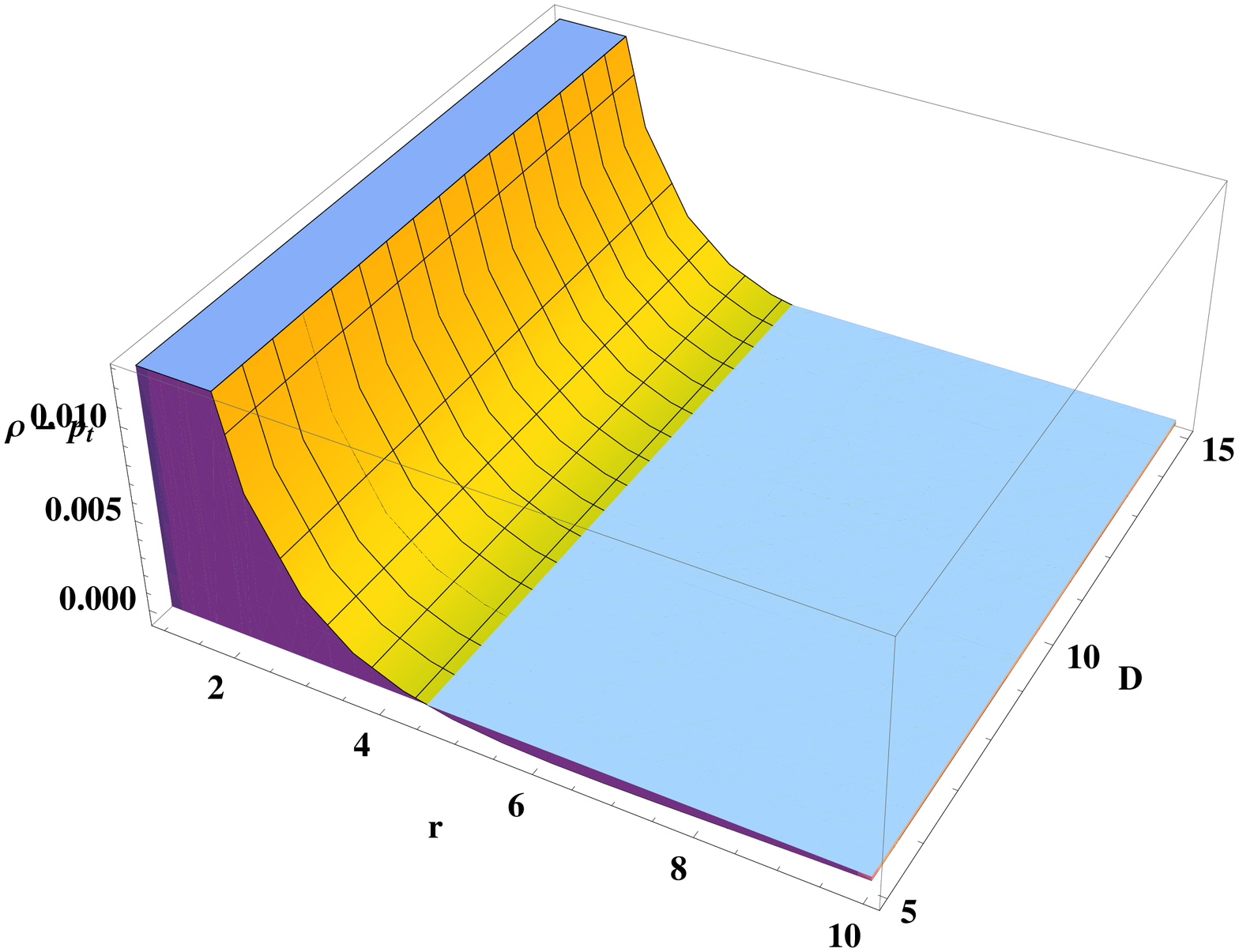}
\includegraphics[width=0.50\textwidth]{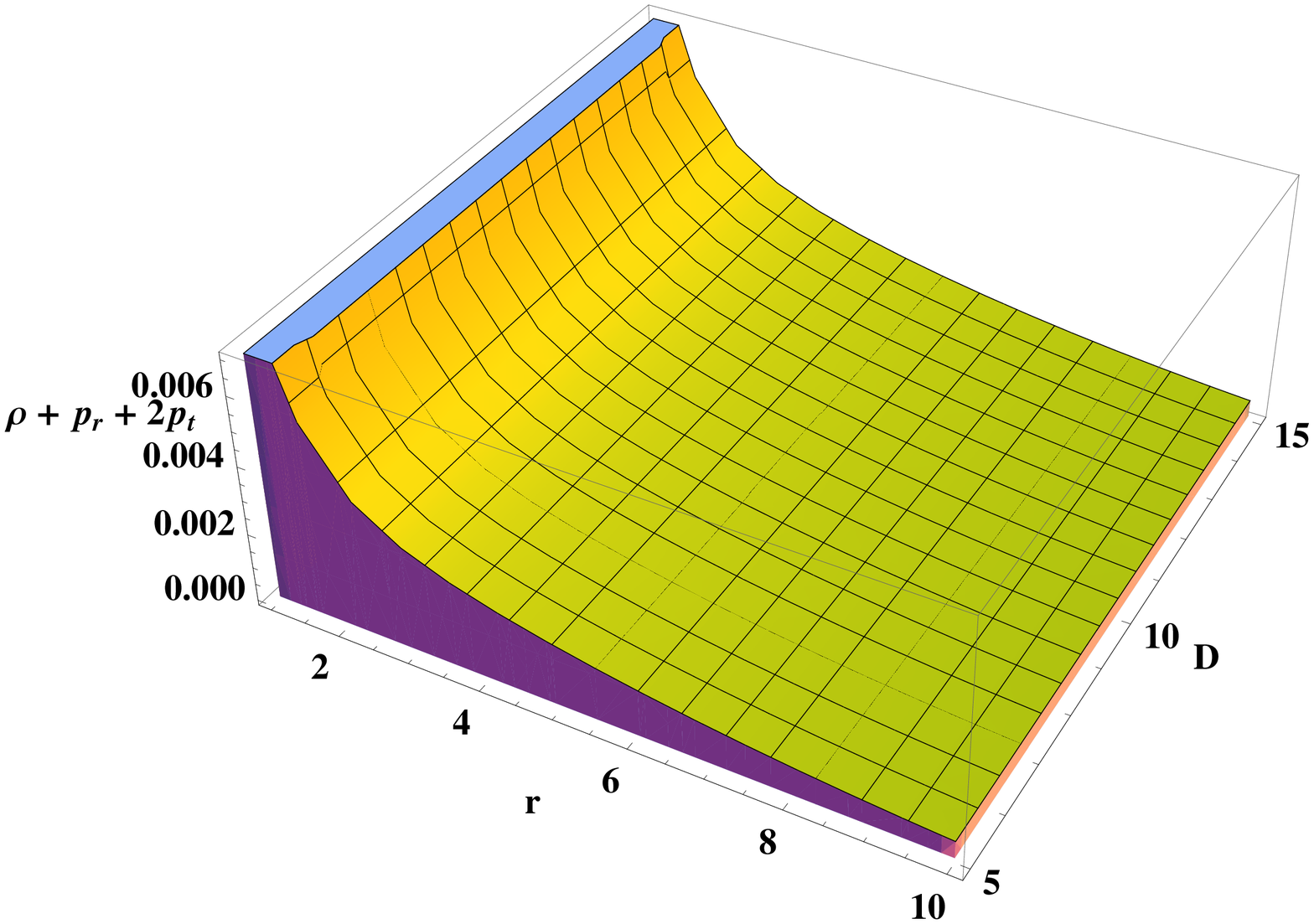}
\end{tabular}
\caption{Validation of energy conditions of singularity-free compact star}
\label{fig:4.jpg}
\end{figure*}

From Fig. 4 below, we observe that the all energy conditions valid for radial pressure as well as tangential pressure with certain range of D. So the compact star presented in this paper is composed of non-exotic matter. Moraes and Sahoo \citep{Moraes/2017} have also constructed the model of wormholes composed by non-exotic matter in the trace of energy momentum-tensor squared gravity. Further it is interesting to note that we may avoid the presence of exotic matter in the framework of $f(R,T)$ gravity and hence no candidate of dark energy/matter is required to explain the accelerating feature of universe as reported in Refs. \citep{Moraes/2017,Yadav/2018a}.\\

The value of $A$ is constraint by employing the energy condition at the center. i.e.\\
(i)NEC:~ $(\rho)_{0} \geq 0$ $\Rightarrow$ $A \geq 0$ \\
(ii)WEC:~ $(\rho)_{0} + (p_{r})_{0} \geq 0$ and $(\rho)_{0} + (p_{t})_{0} \geq 0$ $\Rightarrow$ $A+\alpha A \geq 0$ \&$A+\beta A \geq 0$\\
(iii)DEC:~ $(\rho)_{0} - (p_{r})_{0} \geq 0$ and $(\rho)_{0} - (p_{t})_{0} \geq 0$ $\Rightarrow$ $A - \alpha A \geq 0$ \&$A - \beta A \geq 0$\\
(iv) SEC:~ $(\rho)_{0} + (p_{r})_{0} \geq 0$ and $(\rho)_{0} + (p_{r})_{0} + 2(p_{t})_{0} \geq 0$ $\Rightarrow$ $A+\alpha A \geq 0$ \&$A+(\alpha + 2\beta A) \geq 0$\\

In general theory of relativity, the stellar objects with violations of energy conditions are common. So, there are variety of toy models of stellar objects in which the matter source is in the form of Chaplygin gas \citep{Rahaman/2008}. But in this paper, we have constructed the model of compact star with in the $f(R,T)$ formalism that validate all energy conditions and thus represents a viable model of compact star. The value of $A$ is restricted by equation (\ref{A}).\\

\subsection{Stability}
For Physically acceptable model, the velocity of sound should be less than the velocity of light i.e. $0 \leq v_{s} \leq 1$.\\
\begin{equation}
\label{sr}
v_{sr}^{2} = \frac{dp_{r}}{d\rho} = \alpha
\end{equation}
\begin{equation}
\label{st}
v_{st}^{2} = \frac{dp_{t}}{d\rho} = \beta
\end{equation}
Since both $\alpha$ and $\beta$ lie in between 0 and 1 $(0\leq \alpha \leq 1; 0\leq \beta \leq 1 )$ which implies that velocity of sound is less than 1. Thus our solution validate the existence of physically viable compact star with in the specification of alternative theory of gravity.\\
Equations (\ref{sr}) and (\ref{st}) lead to
\begin{equation}
\label{st-sr}
\mid v_{st}^{2} - v_{sr}^{2}\mid = \mid \beta - \alpha \mid \leq 1
\end{equation}
From equation (\ref{st-sr}), the stability of compact star depends upon the free parameter $\alpha$ and $\beta$. According to Herrera \citep{Herrera/1992}, the region of stellar object in which the radial speed of sound is greater than the transverse speed of sound, is a potentially stable region. Thus, by imposing restriction on the values of $\alpha$ and $\beta$, one may check the stability of derived model.\\
\begin{figure}[tbp]
\begin{center}
\includegraphics[width=0.5\textwidth]{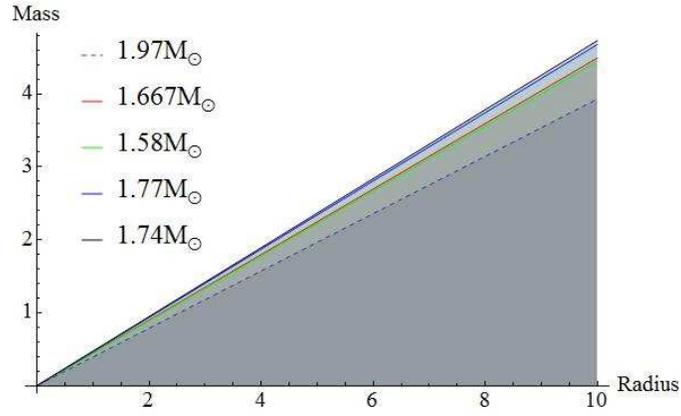}
\caption{Profile of  Mass versus Radius.}
\label{fig:1.jpg}
\end{center}
\end{figure}
\begin{table*}
\small
\caption{Comparision of estimated value of model parameters with observed data sets.\label{tbl-2}}
\begin{tabular}{@{}crrrrrrrrrrr@{}}
\tableline
S. N. & Compact star & $M_{Obs}~(M_\odot)$ & Radii $(r_{\odot})$ & $\zeta$ & $M_{Esti}$ & $Z_{Obs}$ & $Z_{Esti}$\\
\tableline
1. & PSRJ 1614-2230 & $1.97\pm 0.04$~\cite{Demo/2010} & $13\pm 2$ & $6$ & 1.973 & 0.344793 & 0.345475\\
2. & PSRJ 1903+327 &  $1.667\pm 0.02$~\cite{Gang/2013} & $9.438\pm 0.03$ & $4$ & 1.686 & 0.444945 & 0.407709\\
3. & 4U 1820-30 &  $1.58\pm 0.06$~\cite{Guver/2010} & $9.1\pm 0.4$ & $4$ & 1.592 & 0.431786 & 0.393753\\
4. & VelaX-1 &  $1.77\pm 0.08$~\cite{Gang/2013} & $9.56\pm 0.08$ & $3.65$ & 1.814 & 0.484428 & 0.441777\\
5. & 4U 1608-51 &  $1.74\pm 0.14$~\cite{Guver/2010a} & $9.3\pm 1.0$ & $3.65$ & 1.739 & 0.493929 & 0.489662\\
\tableline
\end{tabular}
\end{table*}
\subsection{Adiabatic index}
The adiabatic index is read as
\begin{equation}
\label{ad}
\Gamma = \left(\frac{\rho + p_{r}}{p_{r}}\right)\frac{dp_{r}}{d\rho} = 1+\alpha
\end{equation}
For stable configuration $\Gamma$ should be greater than 1.33 with in the isotropic stellar system. Note that $\Gamma = 1.33$ is the critical value as reported in Refs. \citep{Chandrasekhar/1964,Bondi/1964}. {Equation (\ref{ad}) gives clue to choose the value of free parameter $\alpha$. For stable configuration, we have to choose $\alpha \geq 0.33$ that is why in this paper, we have choosen $\alpha = 0.4$ for graphical ( see fig. 6) and numerical analysis of the model.}\\
\begin{figure}[tbp]
\begin{center}
\includegraphics[width=0.5\textwidth]{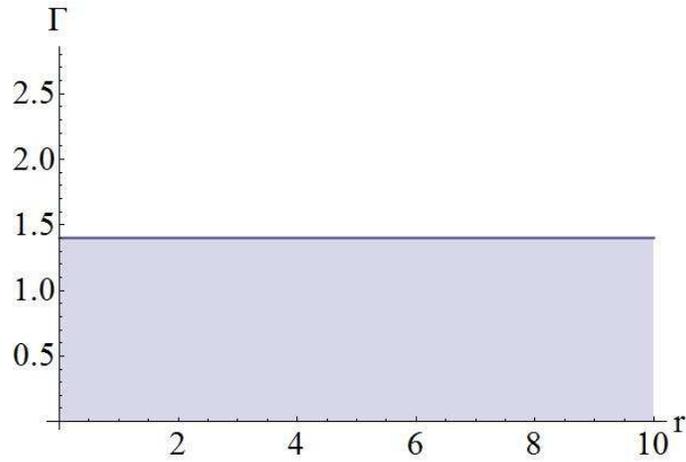}
\caption{Profile of  $\Gamma$ versus $r$.}
\label{fig:1.jpg}
\end{center}
\end{figure}
\subsection{Mass-radius relation}
In our model, the gravitational mass $m(r)$ in terms of radius $r$ is expressed as
\begin{equation}
\label{mass}
m(r) =\int_{0}^{r}4\pi r^{2}\rho dr = \frac{4\pi r[exp(-Ar^{2})(2Ar^{2}-1)+1]}{8\pi + \zeta - \zeta(\alpha+2\beta)}
\end{equation}
The profile of mass function $m(r)$ with respect to radius r for different values of $\zeta$ is shown in Fig. 5.\\
At $r = R$, the gravitational mass is read as
\begin{equation}
\label{mass-1}
m(r)_{r =R} = \frac{4\pi r[(2AR^{2}-1)\left(1-\frac{2M}{R}\right)+1]}{8\pi+\zeta-\zeta(\alpha+2\beta)}
\end{equation}

\begin{figure}[tbp]
\begin{center}
\includegraphics[width=0.5\textwidth]{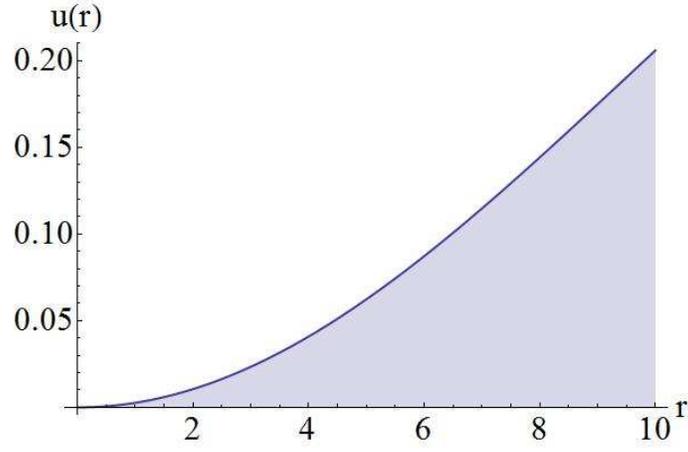}
\caption{Profile of  $u(r)$ versus $r$.}
\label{fig:1.jpg}
\end{center}
\end{figure}
\subsection{Compactness and red-shift}
The compactness of star $(u(r))$ is read as
\[
u(r) = \frac{m(r)}{r}
\]
\begin{equation}
\label{compactness}
 = \frac{4\pi [exp(-Ar^{2})(2Ar^{2}-1)+1]}{8\pi + \zeta - \zeta(\alpha+2\beta)}
\end{equation}
The profile of compactness of star with respect to r is graphed in Figure 7.\\
Therefore, the red-shift function $Z(r)$ is computed as
\begin{equation}
\label{redshift}
Z(r) = (1-2u)^{-\frac{1}{2}}-1 = \left[1-\frac{8\pi [exp(-Ar^{2})(2Ar^{2}-1)+1]}{8\pi + \zeta - \zeta(\alpha+2\beta)}\right]^{-\frac{1}{2}}-1
\end{equation}
The profile of red-shift function with respect to r is depicted in Figure 8.\\
\begin{figure}[tbp]
\begin{center}
\includegraphics[width=0.5\textwidth]{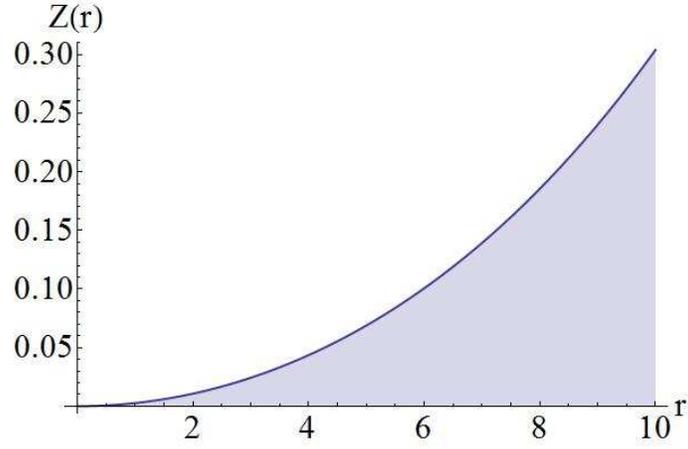}
\caption{Profile of  $Z(r)$ versus $r$.}
\label{fig:1.jpg}
\end{center}
\end{figure}
\section{Physical Validity of Model}
In this subsection, we match the similarity of physical parameters of derived model with their observational values for certain choice of $\zeta$. By using the observational data sets for mass $(M_{\odot})$ and radii $(r_{\odot})$, we carry out a comparative study of estimated mass $(M_{Esti})$, observed red-shift $(Z_{Obs})$ of derived model with observed mass and red-shift of different stars namely PSRJ1614-2230, PSRJ1903+327, 4U1820-30, VelaX-1 and 4U1608-52 and the results are listed in Table 2. From Table 2, we observe that the derived model is very close to 4U 1608051 and PSRJ 1614-2230 for $\zeta = 3.65$ and $\zeta = 6$ respectively. Note that all the figures have been graphed for $\zeta = 3.65$.\\
\section{Result and Discussion}
We have constructed, in the present paper, a singularity-free anisotropic compact star in the framework of $f(R,T)$ gravity. The exact and singularity-free solution of gravitationally collapsing system is obtained by taking into account the well known equations of state which give the relation between energy density and pressure. The energy density, radial pressure and tangential pressure are decreasing function of r. At the center of compact star, $\rho$, $p_{r}$ and $p_{t}$ have certain fixed values which satisfy the relations $\sim$ $(p_{r})_{0} = \alpha(\rho)_{0}$ and $(p_{t})_{0} = \beta(\rho)_{0}$. The behavior of anisotropic parameter has been graphed in Fig. 3 for the two different choice of $\alpha$ and $\beta$. Indeed, the anisotropy in stellar object representing a force which will directed outward when $p_{t} > p_{r}$ and inward if $p_{t} > p_{r}$ which allow the construction of more or less massive distribution respectively \citep{Rahaman/2012}. From Table 1, we observe that the estimated mass of derived compact star  is good agreement with observed mass data sets.  \citep{Demo/2010,Guver/2010,Gang/2013,Guver/2010a}. In general, our solution validate all the energy conditions throughout the stellar region of compact star. The validation of energy conditions can be check by the Fig. 4 above, which is graphed by taking $A=0.025$. Equation (\ref{st-sr}) exhibits the stability criteria of compact star which shows that the particular choice of $\alpha$ and $\beta$ will generates the stable compact star. Let us now concentrate on the some other models of stellar objects with in $f(R,T)$ formalism, especially the work by Moraes and Sahoo \citep{Moraes/2017} and Das et al \cite{Das/2017}. In Refs. \citep{Das/2017}, authors have proposed unique model of stellar object in $f(R,T)$ theory of gravity and show that the gravastar is a viable alternative of black hole. It is worth to note that mechanism of obtaining solution is entirely different from the mechanism adopted in Refs. \cite{Das/2016}. \\

As a final comment, we note that the present study represents the model of non-exotic compact star which validate the SEC as well as other energy conditions in the stellar region of compact star as a significance of extra-term of the $f(R,T)$ theory namely $\zeta T$. The T-dependence of the $f(R,T)$ theory may characterize in describing the physical facts, which is missing in general theory of relativity. In our previous work \citep{Rahaman/2012}, we have investigated a singularity free dark energy star which contains an anisotropic matter that is confined within the certain radius from from center while in the present work, we propose a model of singularity free non exotic compact star with aid of $f(R,T)$ theory of gravitation. In future, one can check the viability of such solution under the specification of other valuable functional forms of $f(R,T)$ such as $f(R,T) = R + \zeta RT $ and
$f(R,T) = R+\zeta R^{2}+\lambda T $ where $\zeta$ and $\lambda$ are arbitrary constants.\\

 \section*{Acknowledgement}
FR  would like to thank the authorities of the Inter-University Centre for Astronomy
and Astrophysics, Pune, India for providing the research facilities. FR is  also
thankful to DST-SERB, Govt. of India and RUSA 2.0, Jadavpur University  for financial support. We are grateful to the referee and Editor for several fruitful suggestions to improve the paper.\\



\begin{thebibliography}{000}
\bibitem[\protect\citeauthoryear{Harko et al.}{2011}]{Harko/2011} Harko, T., Lobo,  F. S. N., Nojiri,  S. and Odintsov, S. D.  , Phys. Rev. D ,{84}, 024020 (2011)

\bibitem[\protect\citeauthoryear{Riess et al.}{2004}]{Riess/2004} Riess et al. A. G. , Astrophys. J. {607}, 665 (2004)

\bibitem[\protect\citeauthoryear{Akerib et al}{2017}]{Akerib/2017} Akerib et al. D. S. , {Phys. Rev. Lett.} {118}, 021303 (2017)

\bibitem[\protect\citeauthoryear{Antoniadis et al.}{2013}]{Antoniadis/2013} Antoniadis et al. J. , {Science} {340}, 1233232 (2013)

\bibitem[\protect\citeauthoryear{Yadav}{2018}]{Yadav/2018a} Yadav, A. K. , Braz. J. Phys. \textbf{49}, 262 (2019)

\bibitem[\protect\citeauthoryear{Moraes and Sahoo}{2017}]{Moraes/2017} Moreas, P. H. R. S. \&  Sahoo, P. K., {Phys. Rev. D} {96}, 044038 ( 2017)

\bibitem[\protect\citeauthoryear{Yadav}{2014}]{Yadav/2014} Yadav A. K. , {Euro Phys. J. Plus} {129}, 194 (2014)

\bibitem[\protect\citeauthoryear{Yadav and Ali}{2018}]{Yadav/2018} Yadav, A. K. \&  Ali A. T. , {Int. J. Geom. Methods in Mod. Phys.} {15}, 1850026 (2018)

\bibitem[\protect\citeauthoryear{Moraes}{2015}]{Moraes/2015} Moreas P. H. R. S. , {Euro. Phys. J. C.} {75}, 168 (2015)

\bibitem[\protect\citeauthoryear{Singh and Kumar}{2014}]{Singh/2014} Singh C. P. \& Kumar P. , {Eur. Phys. J. C.} {74}, 3070 (2014)

\bibitem[\protect\citeauthoryear{Shabani and Farhoudi}{2013}]{Shabani/2013} Shabani H. \& Farhoudi M. , {Phys. Rev. D} {88}, 044048 (2013)

\bibitem[\protect\citeauthoryear{Shabani and Farhoudi}{2014}]{Shabani/2014} Shabani H. \& Farhoudi M. , {Phys. Rev. D} {90}, 044031 (2014)

\bibitem[\protect\citeauthoryear{Sarif and Zubair}{2014}]{Sharif/2014} Sharif  M. \& Zubair  M. , {Astrophys. Space Sc.} {349}, 457 (2014)

\bibitem[\protect\citeauthoryear{Reddy and Kumar}{2013}]{Reddy/2013} Reddy D. R. K. \& Kumar R. S. , {Astrophys. Space Sc.} {344}, 253 (2013)

\bibitem[\protect\citeauthoryear{Bowers and Liang}{1974}]{Bowers/1974} Bowers R. L. \& Liang, {J. Astrophys.} {188}, 657 ( 1974)

\bibitem[\protect\citeauthoryear{Herrera et al.}{2004}]{Herrera/2004} Herrera L., Prisco  A. Di \&  Martin J. et al., {Phys. Rev. D} {69} 084026 ( 2004)

\bibitem[\protect\citeauthoryear{Sharma et al.}{2001}]{Sharma/2001} Sharma R. Mukherjee  S. \& Maharaj S. D., {Gen. Relativ. Gravit.} {33} 999 ( 2001)

\bibitem[\protect\citeauthoryear{Aqueros et al.}{2006}]{Aqueros/2006} Agueros M. A. et al. , Candidate Isolated Neutron Stars and Other Optically Blank X-Ray Fields Identified from the ROSAT All-Sky and Sloan Digital Sky Surveys ; { AJ} {131}, 1740-1749 (2006)

\bibitem[\protect\citeauthoryear{Hewish et al.}{1968}]{Hewish/1968} Hewish A. et al. , Observation of a Rapidly Pulsating Radio Source;{Nature} {217} 709(1968)

\bibitem[\protect\citeauthoryear{Maurya et al.}{2015}]{Maurya/2015} Maurya S.K. Gupta Y. K., Ray S. \&  Dayanandan B. , {Eur. Phys. J. C},{75} 225 (2015)

\bibitem[\protect\citeauthoryear{Maurya et al.}{2016}]{Maurya/2016} Maurya S. K., Gupta Y. K., Ray S. \&  Deb Debabrata , {Eur.Phys.J.C} {76} 693 (2016)

\bibitem[\protect\citeauthoryear{Maurya et al}{2017}]{Maurya/2017} Maurya S. K., Gupta  Y.K., Rahaman F., Rahamam  M. \& Banerjee A. , {Annals of Phys.} {385}, 532 (2017)

\bibitem[\protect\citeauthoryear{Das et al}{2016}]{Das/2016} Das A., Rahaman F., Guha B. K. \&  Ray S,  {Eur.Phys.J.C} {76}, 654 (2016)

\bibitem{Paul/2019} Paul A., Majumdar D. \& Modak K. P.,  Pramana  J. Phys. 92, 44 ( 2019)


\bibitem[\protect\citeauthoryear{Maurya et al}{2016a}]{Maurya/2016a} Maurya S.K., Gupta Y.K., Dayanandan Baiju \& Ray Saibal , {Eur.Phys.J.C} {76}, 266 (2016)

\bibitem[\protect\citeauthoryear{Aziz et al}{2016}]{Aziz/2016} Aziz Abdul, Ray Saibal and Rahaman Farook , {Eur.Phys.J. C} {76} 248 (2016)

\bibitem[\protect\citeauthoryear{Rahaman et al}{2014}]{Rahaman/2014a} Rahaman Farook et al. , {Eur.Phys.J. C} {74} 3126 (2014)

\bibitem[\protect\citeauthoryear{Momeni et al}{2017}]{Momeni/2017} Momeni D., Faizal M., Myrzakulov K. \& Myrzakulov R. , {Eur.Phys.J. C} {77}, 37 (2017)

\bibitem[\protect\citeauthoryear{Moraes et al}{2018}]{Moraes/2018} Moreas, P. H. R. S. \&  Sahoo P. K. , {Phys. Rev. D} {97}, 024007 (2018)

\bibitem[\protect\citeauthoryear{Zubair et al}{2016a}]{Zubair/2016a} Zubair M. Abbas  G. \&  Noureen I. , {Astrophys. Space Sc.} {361}, 1 (2016)

\bibitem[\protect\citeauthoryear{Alhamazawi et al}{2016}]{Alhamazawi/2016}  Alhamazawi A. \&  Alhamazawi R. , {Int. J. Mod. Phys. D} {25}, 1650020 (2016)

\bibitem[\protect\citeauthoryear{Rahaman et al}{2014}]{Rahaman/2014} Rahaman F. et al. , {Int. J. Theor. Physics} {53}, 1910 (2014)

\bibitem[\protect\citeauthoryear{Zubair et al}{2016}]{Zubair/2016} Zubair M., Waheed S. \& Ahmad Y. , {Euro. Phys. J. C.} {76}, 444 (2016)

\bibitem[\protect\citeauthoryear{Krori et al}{1975}]{Krori/1975} Krori K. D. \& J. Barua, {J. Phys. A: Math. Gen} {8}, 508 ( 1975)

\bibitem[\protect\citeauthoryear{Rahaman et al}{2012}]{Rahaman/2012} Rahaman F. et al., {Gen. Relat. Grav.} {44}, 107 ( 2012)

\bibitem[\protect\citeauthoryear{Azreg}{2015}]{Azreg/2015} Azreg-Ainou M. , {J. Cosmol, Astropart. Phys.} {07}, 037 (2015)

\bibitem[\protect\citeauthoryear{Buchdahl}{1959}]{Buchdahl/1959} Buchdahl H. A. , {Phys. Rev.} {116}, 1027 (1959)

\bibitem[\protect\citeauthoryear{Guver et al}{2010a}]{Guver/2010a} Guver T., Wroblesski P., Camarota L.,  Ozel F. , {Astrophys. J.} {712}, 964 (2010)

\bibitem[\protect\citeauthoryear{Hochberg and Visser}{1998}]{Hochberg/1998} Hochberg D. \& Visser M., {Phys. Rev. D} {58}, 044021 ( 1998)

\bibitem[\protect\citeauthoryear{Rahaman et al}{2008}]{Rahaman/2008} Rahaman F., Kalam M., \& Rahaman K. A. , {Mod. Phys. Lett. A} {23}, 1199 (2008)

\bibitem[\protect\citeauthoryear{Herrera}{1992}]{Herrera/1992} Herrera L. , {Phys. Lett. A} {165}, 206 (1992)

\bibitem[\protect\citeauthoryear{Chandrasekhar}{1964}]{Chandrasekhar/1964} Chandrasekhar S. , {Astrophys. J.} {140}, 417 (1964)

\bibitem[\protect\citeauthoryear{Bondi}{1964}]{Bondi/1964} Bondi H. , {Proc. R. Soc. Lond. Series A Math. Phys. Sci} {281}, 39 (1964)

\bibitem[\protect\citeauthoryear{Demorest et al}{2010}]{Demo/2010} Demorest P. B., Pennucci  T.,  Ransom S. M., Roberts M.S.E., Hessels J. W. T.  , {Nature} {467}, 1081 ( 2010)

\bibitem[\protect\citeauthoryear{Guver et al}{2010}]{Guver/2010} Guver T., Wroblesski P., Camarota L.,  Ozel F., {Astrophys. J.} {719}, 1807 ( 2010)

\bibitem[\protect\citeauthoryear{Gangopadhyay et al}{2013}]{Gang/2013} Gangopadhyay T., Ray  S., Li  X. D., Dey J., Dey M. , {Mon. Not. R. Astron. Soc.} {431}, 3216 (2013)

\bibitem[\protect\citeauthoryear{Das et al}{2017}]{Das/2017} Das A., Ghosh S., Guha B. K., Das S.,  Rahaman F. \&  Ray S., {Phys. Rev. D} {95}, 124011 ( 2017)


\end{thebibliography}
\end{document}